\documentclass[prd,aps,twocolumn,preprintnumbers, showpacs, nofootinbib,superscriptaddress,notitlepage]{revtex4-1}
\usepackage{amssymb,amsthm,amsmath}
\usepackage{textcomp}
\usepackage{color}      
\usepackage{slashed}    
\usepackage{verbatim}
\usepackage[normalem]{ulem} 
\usepackage{hyperref}
\usepackage{soul}
\usepackage{xcolor}
\definecolor{purple}{RGB}{128,0,128}

\usepackage{empheq} 
\usepackage{rotating}   
\usepackage{multirow}   
\newcommand{\Tr}{\operatorname{Tr}}

\usepackage{epsfig}

\begin{document}
	\raggedbottom
	
	\title{ 
		Topological quantization of vector-meson anomalous couplings
	}

	\author{   Chao-Qiang Geng
	}	
	\affiliation{School of Fundamental Physics and Mathematical Sciences, Hangzhou Institute for Advanced Study, UCAS, Hangzhou 310024, China} 
	\author{   Chia-Wei Liu  
	}
	\affiliation{School of Fundamental Physics and Mathematical Sciences, Hangzhou Institute for Advanced Study, UCAS, Hangzhou 310024, China} 
	\author{  Yue-Liang Wu
	}	
	\affiliation{School of Fundamental Physics and Mathematical Sciences, Hangzhou Institute for Advanced Study, UCAS, Hangzhou 310024, China} 
	\affiliation{ Institute of Theoretical Physics, Chinese Academy of Sciences, Beijing 100190, China }
	\affiliation{International Centre for Theoretical Physics Asia-Pacific  (ICTP-AP), Beijing 100190, China}
	\affiliation{Taiji Laboratory for Gravitational Wave Universe (Beijing/Hangzhou), University of Chinese Academy of Sciences (UCAS), Beijing 100049, China}

	\date{\today}

	\begin{abstract}
		We identify an overlooked Wess--Zumino--Witten structure in the
		hidden-local-symmetry~(HLS) formulation of vector mesons. The newly identified term
		generically leads to the topological quantization of the vector-meson
		anomalous couplings. If confirmed experimentally, this structure would expose
		the gauge nature of vector mesons in the anomalous sector and single out HLS
		over matter-field descriptions. The observed success of vector-meson dominance in anomalous interactions can then be explained by topological-action saturation of the odd-intrinsic-parity processes.  Precision measurements of
		$\eta^{(\prime)}\to\pi^+\pi^-\gamma^*$ form factors at BESIII and the Super
		$\tau$-Charm Facility can directly test this saturation picture. 
	\end{abstract}

	\maketitle

	\section{Introduction}	
	
	At low energies, the anomalous interactions of pseudoscalar mesons are encoded
	in the Wess--Zumino--Witten (WZW) action~\cite{Wess:1971yu,Witten:1983tw}.
	Its overall coefficient is fixed by anomaly matching and is quantized so that
	the path integral is defined unambiguously. Moreover, the corresponding
	anomalous couplings are protected from higher-order corrections by the
	Adler--Bardeen theorem~\cite{Adler:1969er,Adler:1969gk}, making the anomalous
	sector a rare case in which a coefficient of chiral perturbation theory
	($\chi$PT)~\cite{Weinberg:1978kz,Gasser:1983yg,Gasser:1984gg,Leutwyler:1993iq}
	is both theoretically controlled and directly tied to short-distance
	physics~\cite{Bai:2024lpq}.

	Beyond the anomaly, most interactions in chiral effective theories are
	parameterized by low-energy constants~(LECs), which encode short-distance QCD
	dynamics and must be determined from experiment or nonperturbative inputs
	\cite{Gasser:1983yg,Gasser:1984gg, Bijnens:2001bb,  Bijnens:2014lea}. 
	Their number proliferates at higher orders~\cite{Bijnens:1994ie,Bijnens:1999hw,Bijnens:1999sh,Ebertshauser:2001nj,Li:2024ghg}, and interpreting individual LECs is often nontrivial. A complementary perspective is therefore provided by resonance realizations of chiral dynamics, in which LECs arise as Wilson coefficients after integrating out heavier states~\cite{Ecker:1988te,Braghin:2021qmu,Braghin:2024lbo}. For vector mesons, the hidden local symmetry (HLS) framework offers a systematic and unified description and leads to nontrivial relations among couplings~\cite{Bando:1984ej,Harada:2003jx,Harada:2011xx,Geng:2025fuc}. These relations reduce the number of independent parameters and are closely connected to the   vector-meson dominance~(VMD)~\cite{Sakurai:1960ju}. 
	
	In this work, we consider the HLS framework supplemented by a WZW-like action that bridges these two viewpoints. We show that the newly introduced interaction is quantized in general, and we study its phenomenological implications for anomalous radiative decays and transition form factors. In Sec.~\ref{2}, we present the formalism of the new action. In Sec.~\ref{3}, we examine the constraints from current data. In Sec.~\ref{4}, we conclude.
	
	\section{
		Formalism 
	}\label{2}
	
	Given a unitary field $u$ satisfying $u^\dagger u = 1$, one can construct the one-form
	$\alpha = -i\,d u\,u^\dagger$.
	The WZW action is a boundary functional defined through an auxiliary five-dimensional filling. We write it as~\cite{Kaymakcalan:1983qq,Fujiwara:1984mp,Wu:1986pr,Chou:1983qy}
	\begin{equation}\label{first}
		\!\!\!\!	\Gamma_5[u;M_u^5]
		= 
		\frac{1}{240 \pi ^2}
		\operatorname{Tr}
		\left( 
		\int_{M_u^5}   \alpha ^5 
		+ 5 \int _ {S^4} {\cal I}_4 ( u  , l,r)   \right) \,,
	\end{equation}
	where $\partial M_u^5=S^4$. The boundary $S^4$ is the one-point compactification of physical spacetime, with the fields vanishing at infinity, whereas $M_u^5$ is only an auxiliary filling used to define the WZW functional.
	The first term is evaluated by extending $u$ into $M_u^5$, and the second term is obtained by gauging the first term.
	The gauge-field dependence is implicit in the transformation of $u$.
	For
	$
	u\to g_l u g_r^\dagger\,,
	$
	we require the gauge fields to transform as
	$
	l  \to g_l l g_l ^\dagger - i d g_l g_l ^\dagger
	$
	and
	$
	r  \to g_r r g_r ^\dagger - i d g_r g_r ^\dagger\,.
	$
	Here $g_l$ and $g_r$ are local gauge transformations.
	The explicit form of ${\cal I}_4$ is well established in the literature and is also given in the end matter of this work.
	
	Note that the action is not invariant under an infinitesimal gauge transformation
	$g_{l,r} \simeq 1 + i \epsilon_{l,r}$~\cite{tHooft:1979rat},
	\begin{equation}
		\label{1.3}
		\!\!\!	\delta \Gamma_5[u;M_u^5]
		\!	=
		\!	\frac{1}{48\pi^2}\!
		\int_{S^4}\!\!
		\operatorname{Tr}\!\left[
		\epsilon_l\left(
		i\, d l^{\,3}
		-2\,(d l)^{2}
		\right)
		\right]
		\!	-\!(l\!\to\! r).
	\end{equation}
In $\chi$PT, the relevant gauge groups are $SU(3)_L \otimes SU(3)_R$,
corresponding to gauging the chiral transformations
$q_{L,R} \to \exp(i \theta_{L,R}^a \lambda_a)\, q_{L,R}$, with
$q=(u,d,s)$ and $\lambda_a$ the Gell--Mann matrices. 
	Matching to QCD, one finds that $N_c \Gamma_5 [ U;M_U^5]$ reproduces the anomaly, where $N_c$ is the number of colors.
	Here $U$ denotes the standard mesonic field in $\chi$PT, transforming as $U\to g_L U g_R^\dagger$, with $L$ and $R$ the corresponding gauge fields.
	
	\begin{figure}[t]
		\centering
		\includegraphics[width=\linewidth]{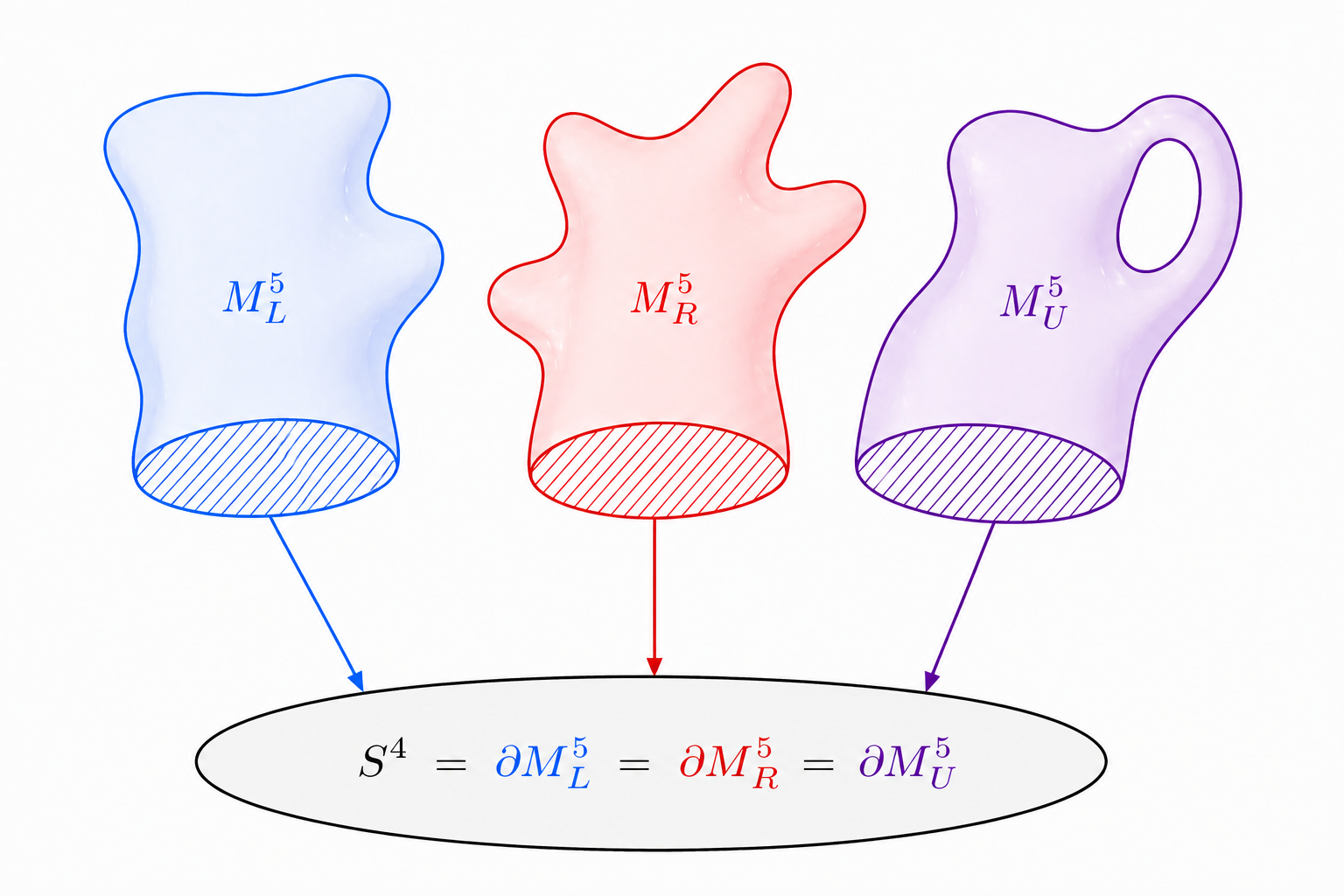}
		\caption{
			Schematic illustration of the independent-filling construction.
			The WZW functionals for ${\color{blue}\xi_L}$, ${\color{red}\xi_R}$, and
			${\color{purple}U}$ are defined by extending the corresponding boundary
			fields into different auxiliary five-manifolds,
			${\color{blue}M_L^5}$, ${\color{red}M_R^5}$, and
			${\color{purple}M_U^5}$, respectively.
			These fillings need not be identified in the bulk, although their
			boundaries are the same physical spacetime,
			$S^4
			=
			{\color{blue}\partial M_L^5}
			=
			{\color{red}\partial M_R^5}
			=
			{\color{purple}\partial M_U^5}$. 
		}
		\label{fig:myfig}
	\end{figure}
	
In the HLS framework, one factorizes the chiral field as
$U=\xi_L^\dagger\xi_R$, with
$\xi_{L,R}\to h\xi_{L,R}g_{L,R}^\dagger$ and
$h(x)\in U_V(3)$; the nonet vector mesons are the corresponding gauge bosons.
The WZW construction defines boundary functionals by extending fields into
auxiliary five-dimensional fillings, while only the induced variation and
filling ambiguity on the physical boundary $S^4$ are physical. Thus, after
introducing the HLS variables, the standard $U$-based WZW functional is not the
only possible structure: $\xi_L$ and $\xi_R$ can also define WZW boundary
functionals with their own auxiliary fillings. We therefore define
\begin{equation}\label{master}
	\begin{aligned}
		\Gamma_{\text{HLS}}
		={}&
		N_h' \Gamma_5[U;M_U^5]
		\\
		&+N_h
		\Big( 
		\Gamma_5[\xi_R;M_R^5]
		-
		\Gamma_5[\xi_L;M_L^5]
		\Big) .  
	\end{aligned}
\end{equation}
Here $N_h$ and $N_h'$ are coefficients. The three fillings have the same
physical boundary,
$\partial M_U^5=\partial M_R^5=\partial M_L^5=S^4$, but in general the
corresponding fields are extended to different five-manifolds, as illustrated
in Fig.~\ref{fig:myfig}.  
	The second term in Eq.~\eqref{master} is a new WZW structure in the HLS formulation.
	In this term, $L$ and $R$ do \emph{not} couple directly, and the relative minus sign is required to keep the HLS symmetry unbroken.
	To preserve parity symmetry, we require $\xi_L\leftrightarrow \xi_R$ under parity.
	Matching to QCD imposes the condition
	$
	N_h' + N_h = N_c .
	$ 
	
	To examine the quantization of the interaction terms, we define
	\begin{equation}
		\begin{aligned}
			w_5[u;S_u^5]
			&=
			\frac{1}{480 \pi ^3}
			\int _{S_u^5} \Tr \left( \alpha ^5 \right) \,, \\
			w_3[u]
			&=
			\frac{-i}{24\pi^2}
			\int_{S^3  } \Tr( \alpha ^3 )\,.
		\end{aligned}
	\end{equation}
If $u$ is globally well defined, then
$w_3[u],\,w_5[u;S_u^5]\in\mathbb{Z}$, since
$\pi_3(SU(3))=\pi_5(SU(3))=\mathbb{Z}$.  
	Here $S^3 $ denotes the one-point compactification of the three-dimensional spatial slice at a fixed time.

	Since the auxiliary filling in Eq.~\eqref{first} is not specified beyond the
	requirement $\partial M_u^5 = S^4$, the exponentiated action must be independent
	of this choice~\cite{Lee:2020ojw,Witten:1983tw}.  For two fillings of the same
	boundary field, $M_u^5$ and $\overline M_u^5$, their difference defines the
	corresponding closed five-manifold
	$S_u^5=M_u^5-\overline M_u^5$, since
	$\partial M_u^5=\partial \overline M_u^5$.  The resulting change of the WZW
	functional is $2\pi$ times the integer winding number on this field-dependent
	closed manifold.
	
	Changing the three fillings in Eq.~\eqref{master} independently therefore
	produces three closed manifolds $S_U^5$, $S_R^5$, and $S_L^5$.  With
	$n_U=w_5[U;S_U^5]$, $n_R=w_5[\xi_R;S_R^5]$, and
	$n_L=w_5[\xi_L;S_L^5]$, one obtains
\begin{equation}\label{qunatized} 
	\Delta\Gamma_{\text{HLS}}
	=
	2\pi
	\left[
	N_h' n_U
	+
	N_h(n_R-n_L)
	\right].
\end{equation} 
Since $n_U,n_R,n_L\in\mathbb{Z}$ are independent, $\exp(i\Gamma_{\text{HLS}})$
is independent of the auxiliary fillings only if $N_h'$ and $N_h$ are
separately quantized, $N_h',N_h\in\mathbb{Z}$.   This separate quantization follows from the independent-filling definition of
	the WZW functionals: the auxiliary extensions of the fields vary
	independently, though they share the fixed physical boundary $S^4$. 
	
The baryon number provides a useful example of a situation in which a common
base space is required. At large $N_c$, it is given by
\begin{equation}
	w_{\bf B}
	=
	\frac{N_h'}{N_c}w_3[U]
	+
	\frac{N_h}{N_c}
	\left(
	w_3[\xi_R]-w_3[\xi_L]
	\right).
\end{equation}
This equation follows from the standard $\chi$PT normalization of the gauged
WZW functional. In particular, varying $N_c\Gamma_5[U;M_U^5]$ with respect to
the background $U(1)_{\bf B}$ gauge field gives the topological baryon number
$w_3[U]$~\cite{Skyrme:1961vq,Witten:1979kh,Witten:1983tx,Karasik:2020zyo}. 
Since all fields entering the baryon number live on the same physical $S^3$,
the Polyakov--Wiegmann identity gives
$w_3[U]=w_3[\xi_R]-w_3[\xi_L]$ and therefore
$w_{\bf B}=w_3[U]$.  This common-base-space condition is specific to the physical spatial slice
and does not apply to the auxiliary five-dimensional fillings in
$\Gamma_{\text{HLS}}$. 
	
	As a formal exercise, one can nevertheless force the three WZW functionals to
	use a common auxiliary filling,
	$M_U^5=M_R^5=M_L^5\equiv M^5$.  In this common-filling model, one obtains
	$n_U=n_R-n_L$ from the Polyakov--Wiegmann identity, so that
	Eq.~\eqref{qunatized} quantizes only the combination $N_h'+N_h$.
	In addition, one has the identity
	\begin{eqnarray}\label{descen}
		&& \Gamma_{\text{HLS}} 
		= 
		N_c	\Gamma_5[U;M^5]
		-
		\frac{N_h }{48\pi^2}
		\int _{S^4}
		{\cal C}_4 \,, \nonumber \\
		&& {\cal C}_4
		={}
		i\alpha_L \alpha_R^3
		+\frac{i}{2}(\alpha_L \alpha_R)^2
		-i\alpha_R \alpha_L^3
		\\
		&& \qquad +
		{\cal I}_4(U,L,R)
		-{\cal I}_4(\xi_R,V,R)
		+
		{\cal I}_4(\xi_L,V,L)\,,\nonumber
	\end{eqnarray}
	with $\alpha_{L,R}=-i\,d\xi_{L,R}\xi_{L,R}^\dagger$.
Even in this restricted case, ${\cal C}_4$ is not a strictly gauge-invariant
local four-form: under a gauge transformation it changes by an exact form,
${\cal C}_4\to{\cal C}_4+d(\cdots)$. 
Thus the restricted common-filling formulation does not reduce the $N_h$ term
to an ordinary gauge-invariant boundary operator. 

	Because the anomalous sector is of order $N_c$, the large-$N_c$ limit suggests that $N_h$ should be proportional to $N_c$.
	In the phenomenological analysis below, we find that the data are well described by $N_h=2N_c$, and hence $N_h'=-N_c$ after anomaly matching.

	\section{
		Numerical results and discussions
	}\label{3}
	
	The four-form ${\cal I}_4$ is the pure transgression form deduced from gauging
	the anomalous five-dimensional WZW action 
	\begin{equation} \label{descent}
		{\cal I}_4(u,l,r)
		=
		\omega_{4,2} \big(l^u, r ,\,0\big)
		-
		\omega_{4,2} \big(l^u ,\, u ^\dagger du ,\,0\big) ,
	\end{equation}
	where $\omega_{4,2}(A_1,A_0,0)$ is the standard transgression 4-form between
	the two connections $A_0$ and $A_1$.  This representative gives the required anomalous variation in Eq.~\eqref{1.3}.
	However, one may still add gauge-invariant terms that contribute to the same
	physical processes, given by 
	\begin{equation}
		{\cal L}_{\rm add}
		=
		\frac{N_c}{16 \pi^2 }
		\sum_{i=1}^4 c_i \Tr \left({\cal L}_i \right) ,
	\end{equation}
where $c_i$ are LECs, and the explicit forms of ${\cal L}_i$ are given in the
end matter and in Refs.~\cite{Harada:2003jx,Fujiwara:1984mp}.

	To examine the effects of the new terms in Eq.~\eqref{master}, 
	we expand the Lagrangian using the convention
	$U = \exp ( 2 i P / f_P )$, where $f_P$ is the meson decay constant.   
	We adopt the same convention for $P$ and $V$ as in our previous
	work~\cite{Geng:2025fuc}, which is also given in the end matter. 
	For the QED case, with $L = R = eQA$ and
	$Q=\mathrm{diag}(2/3,-1/3,-1/3)$, the first few orders in the gauge
	$\xi_L^\dagger = \xi_R=\xi$ are
	\begin{eqnarray}\label{2.4}
		&& \!\!\!\!\!\!\!\!\!
		{\cal L}_{PAA}  =   
		\frac{N_c e^2}{4\pi ^2 f_P }
		\left(    \tilde c_4 -1    
		\right)
		( dA)^2  \operatorname{Tr} \left( Q^2 P \right) \,, \nonumber\\
		&& 	 \!\!\!\!\!\!\!\!\! {\cal L}_{PVV} = 
		\frac{ - N_c }{4 \pi^2 f_P }
		\tilde   c_3 
		\operatorname{Tr}\left(
		( dV) ^2  P 
		\right)\,, \nonumber \\
		&& 	 \!\!\!\!\!\!\!\!\! {\cal L}_{PVA}  = 
		\frac{  N_c e  	
		}{
			8 \pi ^2 f_P 
		}
		( 
		\tilde c_3 - \tilde c_4 
		) 
		\operatorname{Tr}\left( \{
		dV, Q dA 
		\} P \right) \,.   \\ 
		&& \!\!\!\!\!\!\!\!\!	{\cal L}_{PPPA}=
		\frac{ - iN_c e  }{
			12  \pi ^2 f_P ^3 
		} 
		\left[ 
		4 -3 (   \tilde c_{12} + \tilde c_4 ) 
		\right] 
		A  \operatorname{Tr} \left( Q 
		( 	d P )^3 
		\right) \,.\nonumber\\ 
		\label{4po}
		&&	 \!\!\!\!\!\!\!\!\! {\cal L}_{VPPP}  =   \frac{i  N_c  }{4 \pi^2 f_P^3} 
		\left(   \tilde c_3 - \tilde c_{12}   \right) 
		\operatorname{Tr} \left( V (dP)^3 \right) \,, \nonumber 
	\end{eqnarray}
	Only three linearly independent real combinations enter these interactions.
	This reduction follows from the fact that, at low energies after integrating
	out the vector field $V$, the Lagrangians ${\cal L}_{PAA}$ and
	${\cal L}_{PPPA}$ must reproduce the results of conventional $\chi$PT, thereby
	removing two degrees of freedom. The shifted coefficients are
	\begin{equation}\label{eq7}
		\!\!\! 	\left( \tilde c_{12},\,\tilde c_3,\,\tilde c_4 \right)
		\!=\!
		\left(\!c_1\!-c_2\!+\!\frac{N_h}{2N_c},\,
		c_3\!+\!\frac{N_h}{3N_c},\,
		c_4\!+\!\frac{2N_h}{3N_c}\!\right) .
	\end{equation} 
	We stress that the quantities measured in previous experimental studies should be
	identified as $\tilde c_{12}$, $\tilde c_3$, and $\tilde c_4$, since the
	quantized structure $N_h$ was overlooked.
Future theoretical and experimental studies of these anomalous processes should therefore specify whether the fitted quantities are the bare coefficients $c_i$ or the shifted combinations $\tilde c_i$. 
	
	The quantized term with $N_h\neq0$ generates contributions to
	${\cal L}_{PVV}$, ${\cal L}_{PVA}$, and ${\cal L}_{VPPP}$ even when the
	non-quantized coefficients $c_i$ are set to zero.
	We therefore consider the possibility that the quantized HLS contribution is
	dominant. 
	 We refer to this assumption as the
	saturation approximation. In this approximation, the non-quantized
	coefficients are set to $c_i=0$. Using the shifted coefficients defined in
	Eq.~\eqref{eq7}, the saturation value $N_h=2N_c$ gives
	\begin{equation}\label{predict}
		\left(
		\tilde c_{12},\,\tilde c_3,\,\tilde c_4
		\right)_{\rm sat}
		=
		\left(
		1,\,\frac{2}{3},\,\frac{4}{3}
		\right).
	\end{equation}
	For comparison, the conventional VMD pattern corresponds to
	\begin{equation}\label{predict_VMD}
		\left(
		\tilde c_{12},\,\tilde c_3,\,\tilde c_4
		\right)_{\rm VMD}
		=
		\left(\frac{1}{3},\,1,\,1\right).
	\end{equation}
	Both patterns give $\tilde c_3+\tilde c_4=2$, and therefore reproduce the same
	VMD results in channels controlled only by this combination, such as
	$P\to\gamma\gamma^*$ and $\eta\to\pi^+\pi^-\gamma$.  They differ, however, in
	observables sensitive to $\tilde c_4-\tilde c_3$ or to the full $q^2$
	dependence, such as $P\to\gamma^*\gamma^*$ and
	$\eta\to\pi^+\pi^-\gamma^*$, as discussed below.

Before presenting the phenomenological comparison, we emphasize that the
analysis below is intended as a consistency check of the saturation
approximation with $N_h/N_c = 2$, not as a global fit or a precision comparison
with modern state-of-the-art treatments of transition form factors, which
include higher-order and line-shape effects; see
Refs.~\cite{Hanhart:2013vba,Gan:2020aco,Aliberti:2025gmu,Benayoun:2021ody,Holz:2022bkl}
and references therein. The size of chiral corrections is channel dependent
and grows with the typical energy scale of the process. Accordingly, we
organize the discussion by theoretical control, starting from the cleanest
transition-form-factor observables. We do not perform a single global
$\chi^2$ analysis, since it would mix observables with different theoretical
precisions. 
	
	The amplitude for $P \to \gamma(\mu)\,\gamma^*(\nu)$ is given
	by~\cite{Geng:2025fuc,Bernstein:2011bx}
	\begin{equation}
		\Gamma^{\mu\nu}
		=
		\sqrt{2}\,
		\alpha_{\rm em} C_P\,
		\varepsilon^{\mu\nu\alpha\beta}
		q_\alpha p_\beta
		F_{P\gamma}(q^2) ,
	\end{equation}
	where $p$ and $q$ denote the momenta of $P$ and $\gamma^*$, respectively.  The
	coefficient $C_P$ denotes the anomaly-normalized on-shell coupling.  In the
	illustrative HLS implementation used below, the leading corrections to the
	on-shell anomalous amplitude are represented by vector-meson contributions and
	encoded in the form factor $F_{P\gamma}$.  The cleanest channel is
	$\pi^0\to\gamma\gamma^*$, for which the residual correction is expected to be
	of order $(m_\pi/\Lambda_\chi)^4$.  At $q^2=0$, the form factor is normalized
	as $F_{\pi^0\gamma}(0)=1$, with slope
	\begin{equation}\label{new15}
		\frac{\partial F_{\pi^0\gamma}}{\partial q^2}
		=
		\frac{\lambda}{m_\pi^2}
		=
		\frac{\tilde c_3+\tilde c_4}{4}
		\left(
		\frac{1}{m_\rho^2}
		+
		\frac{1}{m_\omega^2}
		\right) .
	\end{equation}
	In the saturation approximation, $\tilde c_3+\tilde c_4=2$, which gives
	\begin{equation}
		\lambda=(3.00\pm0.06)\% ,
	\end{equation}
in good agreement with the experimental value
$\lambda=(3.32\pm0.29)\%$~\cite{ParticleDataGroup:2024cfk}.
Since both the saturation and conventional VMD patterns give
$\tilde c_3+\tilde c_4=2$, this channel tests only the common VMD-like
combination and does not distinguish
$(\tilde c_3,\tilde c_4)=(2/3,4/3)$ from $(1,1)$.

	The decays $\eta^{(\prime)}\to\gamma\gamma^*$ provide complementary tests.
	These transition form factors have been studied extensively in modern
	precision analyses~\cite{Gan:2020aco,Aliberti:2025gmu}. The expected
	theoretical precision of the present $\eta$ estimate is of order
	$(m_\eta/\Lambda_\chi)^4\simeq 6\%$~\cite{Hanhart:2013vba}. The full form
	factors are lengthy and are therefore collected in the end matter. In
	experimental analyses, the $\eta$ form factor is commonly fitted by the dipole form
	\begin{equation}
		F_{\eta\gamma}(q^2)
		=
		\left(
		1-\frac{q^2}{\Lambda^2}
		\right)^{-2}.
	\end{equation}
	Since the corresponding differential data are not publicly available
	yet~\cite{BESIII:2015zpz}, we restrict the comparison to the slope of
	$F_{\eta\gamma}(q^2)$ at $q^2=0$. The saturation approximation gives
	\begin{equation}
		\left[
		\frac{1}{2}
		\partial_{q^2} F_{\eta\gamma}(0)
		\right]^{-1/2}
		=
		(0.76\pm0.01)\ {\rm GeV}.
	\end{equation}
	Compared with the value
	$\Lambda=(0.721\pm0.011)$~GeV~\cite{ParticleDataGroup:2024cfk}, the deviation
	is at the level expected from $(m_\eta/\Lambda_\chi)^4$, consistent with the
	intended theoretical accuracy rather than a precision fit. Performing the same
	estimate for $\eta'\to\gamma\gamma^*$ gives
	\begin{equation}
		\Lambda'=(0.82\pm0.01)\ {\rm GeV},
	\end{equation}
	which agrees well with the experimental value
	$(0.81\pm0.01)$~GeV~\cite{BESIII:2024pxo}. To probe the case with two off-shell
	photons, it would be useful for future experiments to measure the full form
	factors of $\eta\to e^+e^-\mu^+\mu^-$ given in Eqs.~\eqref{form1} and
	\eqref{form2} of the end matter.
	
	A multihadron test of ${\cal L}_{VPPP}$ is provided by
	$\eta^{(\prime)}\to\pi^+\pi^-\gamma^*$~\cite{Stollenwerk:2011zz,Picciotto:1993aa}.
	The corresponding form factor is~\cite{Geng:2025fuc}
	\begin{equation}\label{15}
		\begin{aligned}
			F_V^{[\pi\pi]}
			=&\,
			1
			-
			\frac{3}{4}\tilde c_{12}
			\frac{q^2}{q^2-\bar m_\rho^2}
			-
			\frac{3}{4}\tilde c_4
			\frac{s_\pi}{s_\pi-\bar m_\rho^2}
			\\
			&+
			\frac{3}{4}\tilde c_3
			\frac{
				m_\rho^2(s_\pi+q^2)
			}{
				(s_\pi-\bar m_\rho^2)(q^2-\bar m_\rho^2)
			},
		\end{aligned}
	\end{equation}
	where $\sqrt{s_\pi}$ and $\sqrt{q^2}$ denote the invariant masses of the
	$\pi^+\pi^-$ and $\gamma^*$ systems, respectively.  The effective masses
	$\bar m_V$ include absorptive parts of the propagators~\cite{Zhang:2012gt}.
	A precision treatment of the broad $\rho$ line shape requires analyticity,
	unitarity, and $\pi\pi$ final-state interactions, together with additional
	phenomenological input~\cite{Holz:2022bkl}.  Here we use Eq.~\eqref{15} only
	as a leading description of how the shifted coefficients enter
	$F_V^{[\pi\pi]}$.
	
	For an on-shell photon, $q^2=0$, the form factor $F_V^{[\pi\pi]}$ depends
	only on the combination $\tilde c_3+\tilde c_4$.  The saturation approximation
	gives $\tilde c_3+\tilde c_4=2$, the same value as the conventional VMD
	pattern in Eq.~\eqref{predict_VMD}, and therefore gives a numerically same 
	description of the on-shell $\eta^{(\prime)}\to\pi^+\pi^-\gamma$
	data~\cite{BESIII:2017kyd}.  The distinction between the two patterns appears
	in the off-shell $\gamma^*$ dependence, where $F_V^{[\pi\pi]}$ is sensitive
	separately to $\tilde c_{12}$, $\tilde c_3$, and $\tilde c_4$.  This makes
	$\eta'\to\pi^+\pi^-\mu^+\mu^-$ a useful probe, since the off-shell
	$\gamma^*$ dependence can be measured. Existing experimental analyses, however,
	have implicitly assumed $\tilde c_3=\tilde c_4$~\cite{BESIII:2020elh,BESIII:2024awu}, and the
	differential data are not publicly available.  It would therefore be useful for future analyses to
	determine the full form factor $F_V^{[\pi\pi]}$ in Eq.~\eqref{15}.  For
	$\eta^{(\prime)}\to\pi^+\pi^-e^+e^-$, the partial decay width at high $q^2$
	is suppressed by $m_e^2/q^2$, and the $q^2$ dependence is difficult to resolve
	because of limited statistics~\cite{BESIII:2025cky}.

The measurements discussed above, especially
$\eta^{(\prime)}\to e^+e^-\mu^+\mu^-$ and
$\eta^{(\prime)}\to\pi^+\pi^-\mu^+\mu^-$, would test whether the individual
shifted coefficients follow the saturation pattern. If this pattern is
confirmed experimentally, it would provide a criterion for distinguishing HLS
from ordinary matter-field descriptions of vector mesons. In particular, it
would point to the gauge nature of vector mesons in the anomalous sector, where
the quantized WZW contribution organizes the shifted coefficients $\tilde c_i$.

For pseudoscalar transition form factors, integrating out vector mesons gives
local corrections in powers of $(m_P/\Lambda_\chi)^2$, and VMD assumes that
vector-meson exchange dominates the leading correction. This logic cannot be
directly applied to $\omega$ decays, where the vector-meson kinetic terms must
be retained rather than integrated out. Thus, even under the VMD assumption,
the vector-meson propagator effects need not saturate the
$\mathcal O((m_\omega/\Lambda_\chi)^2)$ corrections. Loop effects at the same
order can be sizable, but they require additional higher-order anomalous
counterterms and LECs. Hence the following comparison of $\omega$ decay
channels should be viewed only as an order-of-magnitude estimate.   
	
	From
	${\cal B}(\omega^0\to\pi^0\gamma)=(8.33\pm0.25)\times10^{-2}$~\cite{ParticleDataGroup:2024cfk},
	and taking $g=5.8\pm0.9$, one obtains
	\begin{equation}
		\tilde c_3+\tilde c_4
		=
		1.95\pm0.30 ,
	\end{equation}
	consistent with the saturation value $\tilde c_3+\tilde c_4=2$. For the
	dilepton modes, the ratios
	\begin{equation}
		R_{\ell\ell}
		\equiv
		\frac{{\cal B}(\omega\to\pi^0\ell^+\ell^-)}
		{{\cal B}(\omega\to\pi^0\gamma)}
	\end{equation}
	remove the overall factor $g^2(\tilde c_3+\tilde c_4)^2$. The saturation
	approximation gives
	\begin{equation}
		R_{ee}^{\rm sat}=8.88\times10^{-3},
		\qquad
		R_{\mu\mu}^{\rm sat}=7.76\times10^{-4},
	\end{equation}
	to be compared with
	$R_{ee}^{\rm exp}=(9.24\pm0.77)\times10^{-3}$ and
	$R_{\mu\mu}^{\rm exp}=(1.61\pm0.22)\times10^{-3}$~\cite{ParticleDataGroup:2024cfk}.
	Thus the electron channel agrees well, while the muon channel is larger than
	the tree-level saturation value by about $3.8\sigma$ using only the
	experimental ratio error. For $\omega\to\pi^+\pi^-\pi^0$, we find
	\begin{equation}
		\Gamma(\omega^0\to\pi^+\pi^-\pi^0)
		=
		(4.0\pm1.3)\ {\rm MeV},
	\end{equation}
	about half of the experimental value
	$(7.74\pm0.13)\ {\rm MeV}$~\cite{ParticleDataGroup:2024cfk}. These deviations
	are compatible with the expected order
	$(m_\omega/\Lambda_\chi)^2\simeq50\%$ uncertainty of this tree-level estimate.
	
\begin{figure}[t]
		\centering
		\includegraphics[width=0.4\linewidth]{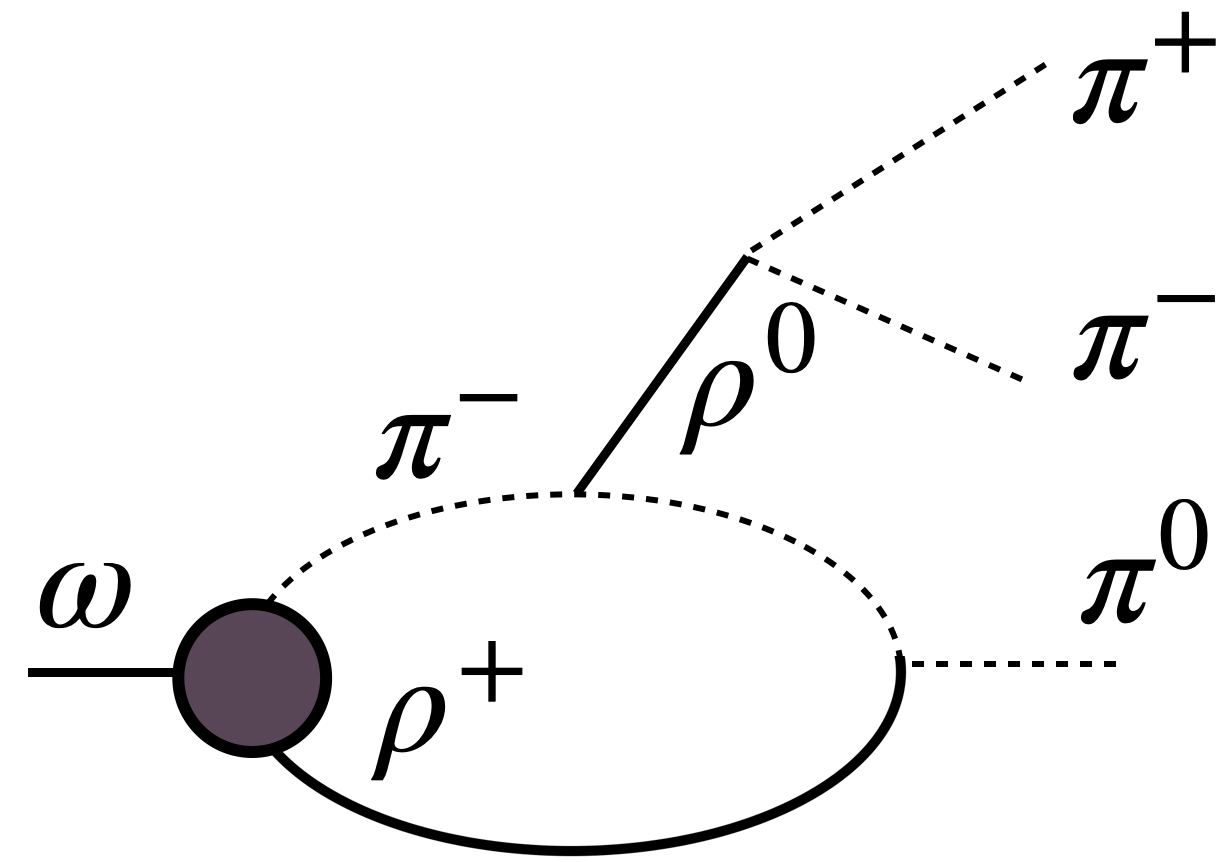}~~~~
		\includegraphics[width=0.4\linewidth]{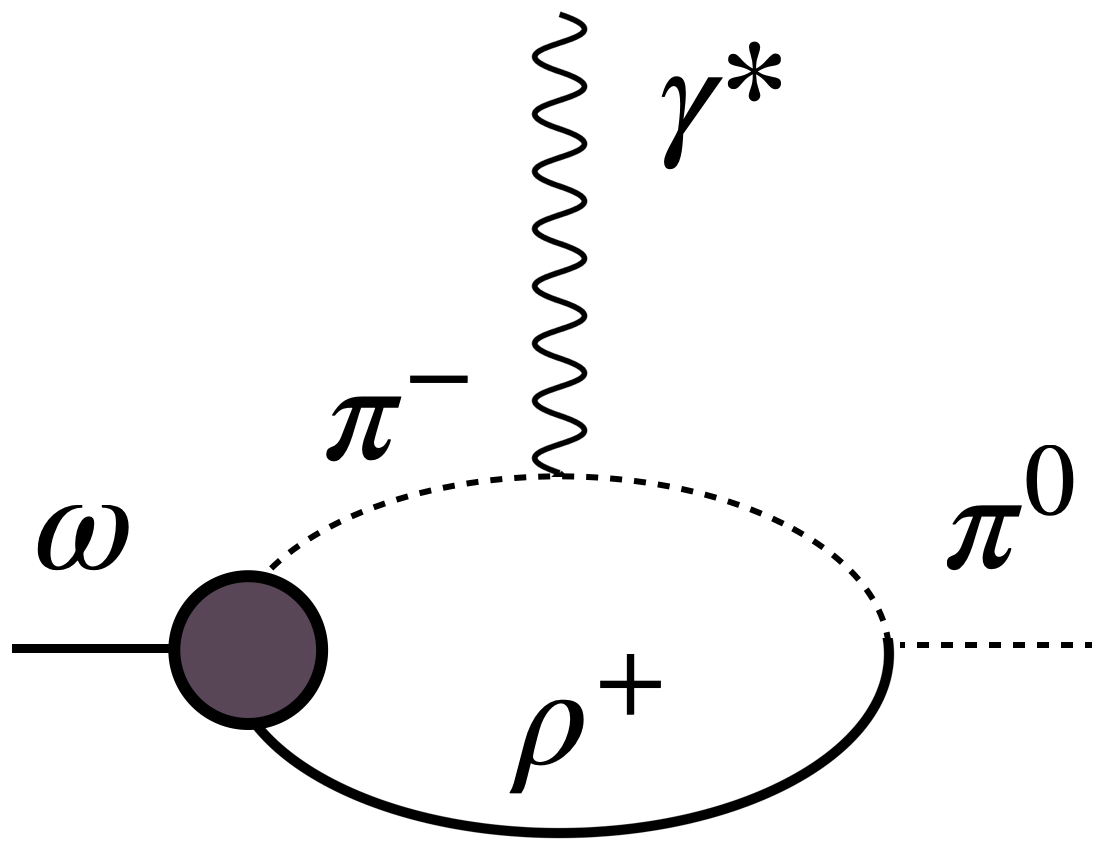}
		\caption{
			Example diagrams contributing to $\omega \to \pi^0\pi^+\pi^-$ and
			$\omega \to \pi^0\gamma^*$ at next-to-leading order are expected to be of
			relative size $(m_V/\Lambda_\chi)^2 \simeq 50\%$.
		}
		\label{fig:omega2pigamma}
	\end{figure}

The enhanced $\omega\to\pi^0\mu^+\mu^-$ ratio and the
$\omega\to\pi^+\pi^-\pi^0$ discrepancy may indicate that non-topological
coefficients $c_i$ are not negligible in $\omega$ channels. However, simply
tuning finite $c_i$ is not a parameter-free resolution: the high-$q^2$ behavior
of $\omega\to\pi^0\mu^+\mu^-$ requires higher-order corrections and refined
line-shape treatments~\cite{Schneider:2012ez}. Representative corrections of
order $(m_\omega/\Lambda_\chi)^2$ are shown in
Fig.~\ref{fig:omega2pigamma}. They affect only the $\omega$-channel saturation
test, not the topological quantization of the HLS WZW contribution proportional
to $N_h$. 
	
	\section{
		Conclusion
	}\label{4}
	In this work, we identified an overlooked WZW  structure in
the HLS formulation of vector mesons. 
The independent auxiliary five-dimensional fillings of the WZW functionals for
$U$, $\xi_R$, and $\xi_L$ reveal that the vector-meson anomalous couplings are
topologically quantized. 
This construction generates the full set of odd-intrinsic-parity vertices and
leads naturally to shifted effective coefficients $\tilde c_i$.  

If confirmed experimentally, this structure would expose the {\it gauge nature}
of vector mesons in the anomalous sector and distinguish HLS from matter-field
descriptions of vector mesons.  As a phenomenological saturation hypothesis,
the shifted coefficients take the pattern in Eq.~\eqref{predict}, which
accounts for part of the success of VMD through destructive interference in
photon-related couplings.  Current data are broadly consistent with this
picture in the better-controlled pseudoscalar channels, while the $\omega$
channels require higher-order effects and should be viewed only as
order-of-magnitude checks.

Future measurements of
$\eta^{(\prime)}\to e^+e^-\mu^+\mu^-$ and
$\eta^{(\prime)}\to\pi^+\pi^-\ell^+\ell^-$ at BESIII and the Super
Tau-Charm Facility~\cite{Achasov:2023gey} can directly test the size of the
quantized anomalous contribution. 
	
	\section*{Acknowledgments} 
	
	This work was supported in part by 
	the National Key Research and Development Program of China under Grant No. 2020YFC2201501;
	the National Science Foundation of China (NSFC) under Grants No. 1821505, 12547104 and 12575096; and the Strategic Priority Research Program and Special Fund of the Chinese Academy of Sciences.


	\clearpage 
	\appendix
	\onecolumngrid
	\begin{center}
		\textbf{\large  End matter  }
	\end{center}

	The gauge-covariant Lagrangian defined in Eq.~\eqref{descent} is
	\begin{equation}
		\begin{aligned}
			{\cal I}_4  = & \, i \left[ 
			\frac{1}{2} ( l \alpha )^2 
			+ l \alpha \, l \, u r u^\dagger 
			+ \frac{1}{4} ( l u r u^\dagger )^2  
			+ ( l^3 - \alpha^2 l ) ( \alpha + u r u^\dagger ) \right]   + d l \, \alpha \, u r u^\dagger
			- (d l \, l + l \, d l) ( \alpha + u r u^\dagger )  \\
			& - i \left[ 
			\frac{1}{2} ( r \beta )^2
			- r \beta \, r \, u^\dagger l \, u
			+ \frac{1}{4} ( r \, u^\dagger l \, u )^2 
			-  \big( r^3 - \beta^2 r \big) ( \beta - u^\dagger l \, u ) \right]   + d r \, \beta \, u^\dagger l \, u
			- ( d r \, r + r \, d r ) ( \beta - u^\dagger l \, u ) \,,
		\end{aligned} 
	\end{equation}  
	where $\beta = u^\dagger \alpha u$. 
	The pseudoscalar and vector meson fields are defined as
	\begin{equation}
		P = 
		\left[ \begin{array}{ccc}
			\frac{\eta_0}{\sqrt{3}} + \frac{\eta_8}{\sqrt{6}} + \frac{\pi^0}{\sqrt{2}} & \pi^{+} & K^{+} \\
			\pi^{-} & \frac{\eta_0}{\sqrt{3}} + \frac{\eta_8}{\sqrt{6}} - \frac{\pi^0}{\sqrt{2}} & K^0 \\
			K^{-} & \bar{K}^0 & \frac{\eta_0}{\sqrt{3}} - \frac{2 \eta_8}{\sqrt{6}}
		\end{array} \right]  \,,~~~
		V  = \frac{g}{\sqrt{2}} \left[  \begin{array}{ccc}
			\frac{1}{\sqrt{2}} (\rho^0 + \omega) & \rho^+ & K^{*+} \\
			\rho^- & \frac{1}{\sqrt{2}} (\omega - \rho^0) & K^{*0} \\
			K^{*-} & \bar{K}^{*0} & \phi
		\end{array} \right]   \,,
	\end{equation} 
	with $g$ the HLS coupling. 
	The HLS Lagrangian that lives on $S^4$ is
	\begin{equation}\label{hlsf}
		{\cal L}  _{S^4} 
		=  
		\frac{f_P^2}{2} \operatorname{Tr}
		\left( \hat{a}_{\perp \mu} \hat{a}_{\perp}^\mu \right)
		+ \frac{m_V^2}{g^2} \operatorname{Tr}
		\left( \hat{a}_{\parallel \mu} \hat{a}_{\parallel}^\mu \right) 
		- \frac{1}{2 g^2}
		\operatorname{Tr} \Big[ 
		( F_V)^{ \mu \nu }  ( F_V) _{ \mu \nu}  
		\Big ] + 
		\frac{N_c}{16 \pi ^2}  
		\sum_{i=1}^4
		c_i  \Tr \left(
		{\cal L}_i \right)
		\,,
	\end{equation}
	where $\hat\alpha_{\parallel,\perp}
	=(\hat\alpha_R\pm\hat\alpha_L)/2$ and
	\begin{equation}
		\hat \alpha _L 
		= -i  
		d \xi _L \xi _L ^\dagger  
		- V 
		+  \xi _L   L \xi ^\dagger  _L \,,~~~
		\hat \alpha _R 
		= -i  
		d \xi _R  \xi _R ^\dagger 
		- V 
		+  \xi _R   R  \xi ^\dagger  _R \,.
	\end{equation}
	The definitions of ${\cal L}_i$ are
	\begin{equation}
		\begin{aligned}
			{\cal L}_1 &= i\left( 
			\hat{\alpha}_L^3 \hat{\alpha}_R - \hat{\alpha}_R^3 \hat{\alpha}_L 
			\right) \,, 
			&
			{\cal L}_2 &= i\left( \hat{\alpha}_L \hat{\alpha}_R \right)^2 \,,
			\\[4pt]
			{\cal L}_3 &= 
			F_V\left( 
			\hat{\alpha}_L \hat{\alpha}_R - \hat{\alpha}_R \hat{\alpha}_L 
			\right) \,,
			&
			{\cal L}_4 &= \frac{1}{2}\left( 
			\xi^\dagger _ L  F_{  L} \xi_L  + \xi _R F_{ R} \xi^\dagger _R
			\right)
			\left( 
			\hat{\alpha}_L \hat{\alpha}_R - \hat{\alpha}_R \hat{\alpha}_L
			\right) \,,
		\end{aligned}
	\end{equation} 
	where $F_{\cal V}=d{\cal V}-i{\cal V}^2$, with ${\cal V}$ denoting any of the
	gauge bosons $V$, $L$, and $R$.
	
	We lay out the amplitudes used in Sec.~\ref{3}.
	For $\eta_8 \to \gamma^*(q_1,\mu)\gamma^*(q_2,\nu)$, we have
	\begin{eqnarray}\label{form1}
		\Gamma^{\mu \nu} &=&e^2 \frac{\sqrt{2} N_c}{12 \pi^2 f_{\eta_8} } \varepsilon^{\mu \nu \alpha \beta} q_{1 \alpha} q_{2 \beta}\left[
		\frac{\sqrt{3}  }{3}
		\left(1-\tilde c_4 
		\right)\right.  
		\nonumber\\
		&& 
		+
		\frac{\sqrt{3} }{4}
		\left(\tilde c_4-\tilde c_3
		\right) 
		\left( 
		D_\rho ( q_1^2) +
		D_\rho ( q_2^2)
		+
		\frac{1}{9}
		\left( 
		D_\omega  ( q_1^2) +
		D_\omega  ( q_2^2)  
		\right) 
		- 
		\frac{4}{9}
		\left( 
		D_\phi  ( q_1^2) +
		D_\phi   ( q_2^2)  
		\right) 
		\right)  \nonumber\\ 
		&&+\frac{\sqrt{3}}{2}
		\tilde c_3  
		\left( D_\rho\left(q_1^2\right) D_\rho \left(q_2^2\right)+
		\frac{1}{9}
		D_\omega \left(q_1^2\right) D_\omega  \left(q_2^2\right)
		-\frac{4}{9}
		D_\phi  \left(q_1^2\right) D_\phi   \left(q_2^2\right)
		\right) \Big] \,.
	\end{eqnarray} 
	while the one for $\eta_0 \to \gamma^*(q_1,\mu)\gamma^*(q_2,\nu)$ is given by
	\begin{eqnarray}\label{form2}
		\Gamma^{\mu \nu} &=&e^2 \frac{\sqrt{2} N_c}{12 \pi^2 f_{ \eta_0}} \varepsilon^{\mu \nu \alpha \beta} q_{1 \alpha} q_{2 \beta}\left[
		\frac{2\sqrt{6}}{3}
		\left(1- \tilde c_4
		\right)\right.  
		\\
		&& 
		+
		\frac{\sqrt{6}}{4}
		\left( \tilde c_4- \tilde c_3
		\right) 
		\left( 
		D_\rho ( q_1^2) +
		D_\rho ( q_2^2)
		+
		\frac{1}{9}
		\left( 
		D_\omega  ( q_1^2) +
		D_\omega  ( q_2^2)  
		\right) 
		+ 
		\frac{2}{9}
		\left( 
		D_\phi  ( q_1^2) +
		D_\phi   ( q_2^2)  
		\right) 
		\right)  \nonumber\\ 
		&&+\frac{\sqrt{6}}{2}
		\tilde	c_3  
		\left( D_\rho\left(q_1^2\right) D_\rho \left(q_2^2\right)+
		\frac{1}{9}
		D_\omega \left(q_1^2\right) D_\omega  \left(q_2^2\right)
		+\frac{2}{9}
		D_\phi  \left(q_1^2\right) D_\phi   \left(q_2^2\right)
		\right)
		\Big] \,.\nonumber 
	\end{eqnarray} 
	Here $D_V(q^2)=m_V^2/(\bar m_V^2-q^2)$. The amplitudes of $\eta$ and $\eta'$
	arise from the mixing between $\eta_8$ and $\eta_0$, given by
	\begin{equation}
		\Gamma^{\mu \nu }_{\eta}
		= \cos\theta_P\,\Gamma^{\mu \nu }_{\eta^8}
		-\sin\theta_P\,\Gamma^{\mu \nu }_{\eta^0}\,,\qquad
		\Gamma^{\mu \nu }_{\eta'}
		= \cos\theta_P\,\Gamma^{\mu \nu }_{\eta^0}
		+\sin\theta_P\,\Gamma^{\mu \nu }_{\eta^8}\,,
	\end{equation}
	where $\theta_P = (-25 \pm 2)^\circ$ is the mixing angle~\cite{Geng:2025fuc}.

\end{document}